\documentclass{article}
\usepackage{latexsym}
\usepackage{graphicx}
\usepackage{epsfig}
\newcommand \be {\begin{equation}}
\newcommand \bea {\begin{eqnarray}}
\newcommand \ee {\end{equation}}
\newcommand \eea {\end{eqnarray}}

\newcommand{\bit}{\begin{itemize}}
\newcommand{\eit}{\end{itemize}}

\newcommand{\eps}{\epsilon}

\begin{document}
\topskip 2cm

\begin{titlepage}

\rightline{\today}

\begin{center}
{\Large\bf Noncommutative Synchrotron} \\

\vspace{2.5cm}

{\large Paolo Castorina$^{1,2,3}$, Alfredo Iorio$^{2,4}$, Dario Zappal\`a$^{3,1}$} \\
\vspace{.5cm}
{\sl $^{1}$ Department of Physics, University of Catania} \\
{\sl Citt\`a Universitaria,  Catania, Italy}\\
{\sl $^{2}$ Center for Theoretical Physics, Massachusetts Institute of Technology}\\
{\sl 77, Massachusetts Avenue, Cambridge MA 02139-4307 U.S.A.} \\
{\sl $^{3}$ Istituto Nazionale di Fisica Nucleare, Sezione di Catania, Italy} \\
{\sl $^{4}$ Istituto Nazionale di Fisica Nucleare, Sede di Presidenza, Italy} \\
\vspace{.5cm}
{E-mail: paolo.castorina@ct.infn.it;  iorio@lns.mit.edu;  zappala@ct.infn.it}\\
\vspace{.5cm}

\vspace{2.5cm}

\begin{abstract}
\noindent We study the departures from the classical synchrotron
radiation due to noncommutativity of coordinates. We find that
these departures are significant, but do not give tight bounds on
the magnitude of the noncommutative parameter. On the other hand,
these results could be used in future investigations in this
direction. We also find an acausal behavior for the
electromagnetic field due to the presence in the theory of two
different speeds of light. This effect naturally arises even if
only $\theta^{12}$ is different from zero.
\end{abstract}
\end{center}

\vspace{2.5cm}



MIT-CTP-3336 $\quad$ hep-th/0212238

\end{titlepage}

\newpage

\section{Introduction}

\noindent The idea of noncommutative space-time coordinates in
physics dates back to the 1940's \cite{snyder}. Recently, due to
the discovery of Seiberg and Witten \cite{sw} of a map (SW map)
that relates noncommutative to commutative gauge theories, there
has been an increasing interest in studying the impact of
noncommutativity on fundamental as well as phenomenological
issues. Two directions seem to us very central: on the one hand,
the clear understanding of the spatiotemporal symmetry and
unitarity properties of these theories \cite{alvarezgaume},
\cite{kostelecky}, \cite{chai}, \cite{doplicher1},
\cite{doplicher2}, \cite{iorio}, \cite{jackiwpi}; on the other
hand the hunting for experimental evidence (see e.g.
\cite{amelia}), possibly in simple theoretical set-ups
\cite{jackiw}.

\noindent The aim of the present paper is to study the effects of
noncommuting space-time coordinates on synchrotron radiation in
{\it classical} electrodynamics. The motivations are twofold: i)
we want to see the fundamental effects of noncommutativity, such
as acausality, and violation of Lorentz and scale invariance,
practically at work in a simple case; ii) we want to explore the
physics of synchrotron radiation in the hope that more stringent
limits on the magnitude of the noncommutative parameter $\theta$
could be set in this case (for a nice account on the current
bounds see for instance \cite{kostelecky}).

\noindent The current views on the space-time properties of
noncommutative field theories are essentially three: i) with the
only exception of space-time translations, spatiotemporal
symmetries are manifestly violated (see for instance
\cite{alvarezgaume}), and sometimes the artifact of the so-called
``observer transformations'' has to be introduced
\cite{kostelecky}, \cite{grosse}; ii) full Lorentz invariance
(including parity and time-reversal) is imposed on (a
dimensionless) $\theta^{\mu \nu}$, leading to a quantum space-time
with the same classical global symmetries \cite{doplicher1}; iii)
being the noncommutative field theory an effective theory of the
fundamental string theory, space-time symmetries are not a big
issue. For a quite comprehensive review see for instance
\cite{review}. We believe that a lot more work is needed to
fully understand this very fascinating matter.

\noindent What we intend to do here is to take a practical view
and tackle the problem of finding corrections to the spectrum of
synchrotron radiation induced by noncommutativity at first order
in $\theta$. We shall find that, in our approximations, these
corrections act as a powerful ``amplifier'' of the effects of
noncommutativity: independently from the actual value of $\theta$
the synchrotron radiation amplifies the effects of a factor
$O(10^{13})$. On the other hand, due to the current bounds on $\theta$ 
this amounts to a correction of $O(10^{-10})$ of the commutative 
counterpart, hence we are still far from possible testable effects. 
We also see in this analysis that some surprising acausal behaviours 
naturally arise even if only the space-space component $\theta^{1 2}$ is 
taken to be different from zero. The effect is due to the presence 
of two different speeds of light in this theory. This result is in
contrast with the general belief that acausality effects should
arise in this context only when $\theta^{0 i} \neq 0$. We take
this last result as a confirmation that the issues of space-time
properties are far from being clarified.

\noindent The theory we shall be dealing with in this paper is
affected with serious problems in the quantum phase (see e.g.
\cite{ruizruiz}, \cite{esperanza}). For instance, the
truncation of the theory at first order in $\theta$ leads to
infrared instabilities at the quantum level \cite{amelia}, and in
the limit $\theta \to 0$ the commutative quantum electrodynamics
is not recovered \cite{seiberg}. These obnoxious features seem to
be related to an unusual correspondence among the ultraviolet and
infrared perturbative regimes of the quantum theory. It is still
unclear whether this correspondence is an artifact of the
perturbative calculations or a more fundamental (hence more
serious) problem. For instance in \cite{vaidya} it is shown that
there are scalar field theories where the connection is actually
absent. These facts evidently mean that the quantum theory is
still a ``work in progress'', and we shall not further address
these important matters here. We shall instead focus on the
classical theory, in the hope that a meaningful quantum theory
might be discovered in the future, and that such a theory could
have the classical model we are about to use as a limit. After all
classical general relativity is widely used and experimentally
tested even if a sound quantum theory of gravity still does not
exist (and may as well not exist at all).

\noindent In the next Section we shall recall the main ingredients
of noncommutative electrodynamics \cite{jackiw}, and set the
notation. In Sections 3 we shall exhibit the electromagnetic
potentials for the noncommutative synchrotron, while in Section 4
we shall give the approximate expressions for the electric and
magnetic fields to estimate the leading corrections to synchrotron
radiation. Finally, in Section 5 we shall draw our conclusions.

\section{Noncommutative Electrodynamics}

\noindent For us the noncommutativity of space-time coordinates
will be expressed in the simplest possible fashion, the canonical
form \cite{wess}, given by
\begin{equation}  \label{1}
x^\mu * x^\nu - x^\nu * x^\mu = i \theta^{\mu \nu} \;,
\end{equation}
where the Moyal-Weyl $*$-product of any two fields $\phi(x)$ and
$\chi(x)$ is defined as
\begin{equation}
(\phi * \chi) (x) \equiv \exp\{ \frac{i}{2} \theta^{\mu \nu}
\partial^x_\mu \partial^y_\nu \} \phi (x) \chi (y)|_{y \to x}  \label{moyal}
\end{equation}
$\theta^{\mu \nu}$ is $c$-number valued, the Greek indices run
from $0$ to $n-1$, and $n$ is the dimension of the space-time.
This approach, of course, does not contemplate all the possible
ways noncommutativity of the coordinates could take place. For
instance, two equally valid, if not more general,  approaches are
the Lie-algebraic and the coordinate-dependent ($q$-deformed)
formulations \cite{wess}, and many other approaches exist.
Nonetheless, the canonical form is surely the most simple and the
basic features of noncommutativity are captured in this model.

\noindent The action for the noncommutative Maxwell theory for
$n=4$ is
\begin{equation}  \label{ncym}
\hat{I} = - \frac{1}{4} \int d^4 x \hat{F}^{\mu \nu} \hat{F}_{\mu
\nu} \;,
\end{equation}
where $\hat{F}_{\mu \nu} = \partial_\mu \hat{A}_\nu - \partial_\nu
\hat{A}_\mu - i [ \hat{A}_\mu , \hat{A}_\nu ]_* $, $\hat{A}_\mu$
can be expressed in terms of a U(1) gauge field $A_\mu$ and of
$\theta^{\mu \nu}$ by means of the SW map \cite{sw}, $\hat{A}_\mu
(A,\theta)$. Note that $\hat{A}_\mu (A,\theta) \to A_\mu$ as
$\theta^{\mu \nu} \to 0$, hence, in that limit, $\hat{F}_{\mu \nu}
\to F_{\mu \nu} = \partial_\mu A_\nu - \partial_\nu A_\mu$.

\noindent Let us now recall some useful results, valid at all
orders in $\theta$. The Noether currents for space-time
transformations (full conformal group) for the noncommutative
electrodynamics described by Eq. (\ref{ncym}) were obtained in
\cite{iorio}
\begin{equation}\label{noether}
J_f^\mu = \Pi^{\mu \nu} \delta_f A_\nu - \hat{\cal L} f^\mu \;,
\end{equation}
where $\hat{I} = \int d^4 x \hat{\cal L}$,  $\Pi^{\mu \nu} =
\delta \hat{\cal L} / \delta \partial_\mu A_\nu$, and for
translations (the only symmetric case) $f^\mu \equiv a^\mu$, where
$a_\mu$ are the infinitesimal parameters. By making use of the
gauge-covariant transformations \cite{jackiwpi} one finds the conserved
energy-momentum tensor \cite{iorio}
\begin{equation}\label{tmunu}
T^{\mu \nu} = \Pi^{\mu \rho} F^\nu_\rho - \eta^{\mu \nu} \hat{\cal
L} \;,
\end{equation}
whose symmetry is, of course, not guaranteed as in the commutative
case ($\Pi^{\mu \nu} = - F^{\mu \nu}$), and the conservation holds
when the equations of motion
\begin{equation}\label{maxwgen}
\partial_\mu \Pi^{\mu \rho} = 0 \;,
\end{equation}
are satisfied.
Thus, in full generality, the conserved Poynting vector is
\begin{equation}\label{poynting}
  \vec{S} = \frac{c}{4 \pi} \vec{D} \times \vec{B} =
  \frac{c}{4 \pi} \vec{E} \times \vec{H} \;,
\end{equation}
where
\begin{equation}\label{dh}
  D^i \equiv \Pi^{i 0} \quad {\rm and} \quad H^i \equiv \frac{1}{2} \eps^{i j
  k} \Pi_{j k} \;.
\end{equation}

\noindent For our purpose it suffices to treat the simplest case
of noncommutative electrodynamics at order $O(\theta)$, coupled to
an external current, described by
\begin{equation}  \label{othetamaxwell}
\hat{I} = - \frac{1}{4} \int d^4 x \; [F^{\mu \nu} F_{\mu \nu}
-\frac{1}{2} \theta^{\alpha \beta} F_{\alpha \beta} F^{\mu \nu}
F_{\mu \nu} + 2 \theta^{\alpha \beta} F_{\alpha \mu} F_{\beta \nu}
F^{\mu \nu}] + J_\mu \hat{A}^\mu \;,
\end{equation}
where we made use of the $O(\theta)$ SW map
\begin{equation}
\hat{A}_{\mu}(A, \theta) = A_{\mu} - \frac{1}{2} \theta^{\alpha
\beta}A_{\alpha}(\partial_{\beta}A_{\mu} + F_{\beta \mu})  \;,
\end{equation}
and of the $*$-product defined in Eq. (\ref{moyal}). From now on
our considerations will be based on such a $O(\theta)$ theory.

\noindent In Ref. \cite{jackiw} it was found that, in the presence
of a background magnetic field $\vec{b}$, and in absence of
external sources ($J_\mu = 0$), the $O(\theta)$ plane-wave
solutions exist. The waves propagating transversely to $\vec{b}$
travel at the modified speed $c'=c(1 - \vec{\theta}_T \cdot
\vec{b}_T)$ (where $\vec{\theta} \equiv (\theta^{1}, \theta^{2},
\theta^{3})$, with $\theta^{i j} = \eps^{i j k} \theta^{k}$, and
$\theta^{0 i} = 0$) while the ones propagating along the direction
of $\vec{b}$ still travel at the usual speed of light $c$.

\noindent The plane-waves, unfortunately, do not give a stringent
bound on  $\theta$. As a matter of fact, with the current bound of
$10^{-2} ({\rm TeV})^{-2}$ \cite{bound}, one would need a
background magnetic field of the order of 1 Tesla over a distance
of 1 parsec to appreciate the shift of the interference fringes
due to the modified speed of noncommutative light. It is then of
strong interest to find more stringent phenomenological bounds on
the noncommutative parameters. In the next Sections we shall study
the synchrotron radiation in the hope to ameliorate those bounds.

\noindent In order to do that let us recall the {\it linearized} constitutive
relations among the fields following from the modified Maxwell
Lagrangian in Eq. (\ref{othetamaxwell}) \cite{jackiw}
\begin{equation}\label{linearized}
  D^i = \epsilon^{i j} E^j \quad {\rm and} \quad H^i = (\mu^{-1})^{i j}
  B^j \;,
\end{equation}
where
\begin{equation}\label{epsmu}
  \epsilon^{i j} \equiv a \delta^{i j} +  \theta^i b^j + \theta^j b^i \; , \;
  (\mu^{-1})^{i j} \equiv a \delta^{i j} - (\theta^i b^j + \theta^j b^i) \;,
\end{equation}
$a = (1 - \vec{\theta} \cdot \vec{b})$. Since $F_{\mu \nu} =
\partial_{[\mu }A_{\nu]}$ still holds,
\begin{equation}\label{potentials}
\vec{B} = \vec{\nabla} \times \vec{A} \quad {\rm and} \quad
\vec{E} = - \frac{1}{c} \frac{\partial}{\partial t} \vec{A} -
\vec{\nabla} \Phi \;,
\end{equation}
the Bianchi identities are not modified. On the other hand, the
dynamical Maxwell equations (\ref{maxwgen}), when a source is
added, become
\begin{equation}\label{neweleqs}
  \partial_\mu \Pi^{\mu \nu} = J^\nu + \theta^{\alpha \nu}
  J^\sigma \partial_\alpha A_\sigma + \theta^{\alpha \sigma}
  \partial_\sigma (A_\alpha J^\nu) \;,
\end{equation}
leading to (for $\theta^{0 i} = 0$)
\begin{eqnarray}
\vec{\nabla} \cdot \vec{D} & = & 4 \pi [\rho + \vec{\theta} \cdot
\left( \vec{\nabla} \times (\rho \vec{A}) \right)] \;, \label{maxdyn1} \\
\left( \vec{\nabla} \times \vec{H} - \frac{1}{c}
\frac{\partial}{\partial t} \vec{D} \right)^i & = & \frac{4
\pi}{c} \left[ J^i + \theta^j \left( \epsilon^{ijk} J^\sigma
\partial_k A_\sigma + \epsilon^{jlk} \partial_k (A^l J^i) \right)
\right] \label{maxdyn2} \;,
\end{eqnarray}
where the Latin indices run from 1 to 3.

\section{Noncommutative Synchrotron}

\noindent By using the potentials in Eq. (\ref{potentials}), and the Lorentz
gauge $\partial_\mu A^\mu = 0$, the equations of motion
(\ref{maxdyn1}) and (\ref{maxdyn2}) become
\begin{eqnarray}
a \Box \Phi + (\theta^i b^j + \theta^j b^i) \left[ \partial_i
\partial_j \Phi + \frac{1}{c} \frac{\partial}{\partial t}
(\partial_i A_j) \right] = - 4 \pi [\rho + \vec{\theta} \cdot
\left( \vec{\nabla} \times (\rho \vec{A}) \right)] \;, && \label{maxdyn4} \\
a \Box A^i + (\theta^i b^j + \theta^j b^i) \left[ - \frac{1}{c^2}
\frac{\partial^2}{\partial t^2} A_j + \partial_j (\vec{\nabla}
\vec{A}) \right] + \eps^{i k m} (\theta^m b^j + \theta^j b^m)
\eps^{j l p} \partial_k \partial_l A_p & & \nonumber \\
= - \frac{4 \pi}{c} \left[ J^i + \theta^j \left( \epsilon^{ijk}
J^\sigma \partial_k A_\sigma + \epsilon^{jlk} \partial_k (A^l J^i)
\right) \right] \;, &&  \label{maxdyn3}
\end{eqnarray}
where $\Box \equiv -c^{-2}
\partial^2_t + \partial_{x_1}^2 +\partial_{x_2}^2 +\partial_{x_3}^2$.

\noindent For our purpose we use the following settings:
\begin{itemize}
  \item Charged particle moving (circularly) in the plane $(1,2)$
  with speed $c \beta_i (t) = \dot{r}_i (t)$, $i=1,2$, i.e.
\begin{equation}\label{J}
J_\mu = e c \beta_\mu \delta(x_3) \delta^{(2)} (\vec{x} -
\vec{r}(t)) \;,
\end{equation}
  where $\vec{r}(t)$ is the position of the particle, and $\beta_\mu = (1,
  \vec{\beta})$ (hence $J_3 = 0$);
  \item $\vec{b} = (0, 0, b)$, background magnetic field speeding
  up the particle;
  \item $\vec{\theta} = (0, 0, \theta)$, i.e. $\theta^3$ is the only
non-zero component on $\theta^{\mu \nu}$, this is the simplest
possible case to see the effects of noncommutativity.
\end{itemize}
With these settings $\eps^{i j} = a \delta^{i j} + \delta^{i 3}
\delta^{j 3} \lambda$, where $\lambda \equiv 2 \theta b$, $a = (1
- \theta b) = (1 - \lambda / 2)$.

\noindent Furthermore, we are interested in evaluating the larger
noncommutative departures from the synchrotron spectrum, hence
contributions which are of order higher that $O(e/R)$, where $e$
is the electric charge, and $R$ is the distance from the source,
will be neglected. Let us write the general solutions of Eqs.
(\ref{maxdyn4}) and (\ref{maxdyn3}) as $A_\mu = A^{(0)}_\mu +
\theta (A^{(\theta)}_\mu + \bar{A}^{(\theta)}_\mu)$, where
$A^{(0)}_\mu$ is the solution for $\theta = 0$, $A^{(\theta)}_\mu$
is the correction obtained neglecting the $O(\theta)$
contributions to the coupling with the external currents, and
$\bar{A}^{(\theta)}_\mu$ is the correction coming solely from the
$O(\theta)$ contributions to the coupling with the external
currents. It is easy to see that $\bar{A}^{(\theta)}_\mu$ is
$O(e/R)^2$, while $A^{(\theta)}_\mu$ is $O(e/R)$.

\noindent Taking all of the above into account, the approximations
made lead us to write Eq.s (\ref{maxdyn4}), and (\ref{maxdyn3}) as
\begin{eqnarray}
{\Box} {A}_1 + \lambda \partial_2 (\partial_1 {A}_2 -
\partial_2 {A}_1) & = & - \frac{4 \pi}{c} \tilde{J_1} \;, \label{approxa1} \\
{\Box} {A}_2 + \lambda \partial_1 (\partial_2 {A}_1 - \partial_1
{A}_2) & = & - \frac{4 \pi}{c} \tilde{J_2} \;, \label{approxa2} \\
{\Box} {A}_3 - \frac{\lambda}{c^2} \partial^2_t {A}_3 +
\frac{1}{c} \lambda \partial_3 \partial_t
\Phi & = & 0 \;, \label{approxa3} \\
{\Box} {\Phi} + \lambda ( \partial^2_3 \Phi + \frac{1}{c}
\partial_3 \partial_t A_3) & = & - 4 \pi \tilde{\rho} \label{approxphi} \;,
\end{eqnarray}
where $\tilde{J}_i \equiv J_i / a$, $i = 1,2$, and $\tilde{\rho}
\equiv \rho / a$. We notice here that: i) the $1 \leftrightarrow
2$ symmetry for Eq.s (\ref{approxa1}) and (\ref{approxa2}), due to
the rotation symmetry still present on the plane for the
noncommutative case; ii) Eqs. (\ref{approxa1}) and
(\ref{approxa2}) couple the components $A_1$ and $A_2$, while Eqs.
(\ref{approxa3}) and (\ref{approxphi}) couple the components $A_3$
and $\Phi$. When one solves the equations for $A_3$ and $\Phi$,
one sees that $A_3 \sim O(\lambda)$. This gives a negligible
contribution $O(\lambda^2)$ to $\Phi$, but is an effect completely
due to noncommutativity, absent in the standard theory. As a
matter of fact we have $A_3 \neq 0$ even if the current $\vec{J}$
is taken to lay in the plane $(1,2)$.

\noindent Indeed, by writing Eq.s (\ref{approxa1}),
(\ref{approxa2}), (\ref{approxa3}), and (\ref{approxphi}) in the
space of momenta we obtain to order $O(\lambda)$
\begin{eqnarray}
A_1 (\vec{k}, \omega) & = & - \frac{4 \pi}{c} \frac{\tilde{J}_2
(\vec{k}, \omega) \lambda k_1 k_2 + \tilde{J}_1 (\vec{k}, \omega)
[ a (\omega^2 / c^2 - \vec{k}^2) + \lambda k_1^2]} { [ \omega^2 /
c^2 - \vec{k}^2 + \lambda (k_1^2 + k_2^2)]
[a(\omega^2 / c^2 - \vec{k}^2) ]} \;, \label{a1fourier} \\
A_2 (\vec{k}, \omega) & = & A_1 (\vec{k}, \omega) (1 \leftrightarrow 2) \;, \label{a2fourier} \\
A_3 (\vec{k}, \omega) & = & - \lambda \frac{4 \pi \tilde{\rho}
(\vec{k}, \omega) k_3 \omega} { [ \omega^2 / c^2 - \vec{k}^2 +
\lambda \omega^2 ] [a( \omega^2 / c^2 - \vec{k}^2 - \lambda k_3^2) ]} \;, \label{a3fourier} \\
\Phi (\vec{k}, \omega) & = & - \frac{ 4 \pi \tilde{\rho} (\vec{k},
\omega) }{( \omega^2 / c^2 - \vec{k}^2 - \lambda k_3^2)}
\label{phifourier} \;,
\end{eqnarray}
or
\begin{equation}\label{allfourier}
A_\mu (\vec{k}, \omega) \equiv G_{\mu \nu} (\vec{k}, \omega) J_\nu
(\vec{k}, \omega) \;,
\end{equation}
where $\tilde{J}_0 \equiv c \tilde{\rho}$. We can now identify the
Green's functions as
\begin{equation}
G_{\mu \nu} (\vec{R}; \tau) = \frac{1}{4 \pi^3} \int d^3 k d
\omega e^{i [\omega \tau - \vec{k} \cdot \vec{R}]} G_{\mu \nu}
(\vec{k}, \omega) \label{gpl} \;,
\end{equation}
where $\vec{R} \equiv \vec{x} - \vec{x}'$, $\tau \equiv t - t'$,
and we made use of the fact that translation invariance is still
present for the noncommutative theory.

\noindent The non-zero Green's functions are then
\begin{eqnarray}
G_{0 0} (\vec{R}; \tau) & = & \frac{1}{R} \delta(\tau - R / c) -
\lambda \left( \frac{1 - c \tau / R}{R}
\delta(\tau - R / c) + \frac{\tau}{R} \delta'(\tau - R / c) \right) \;, \label{greenphi} \\
G_{1 1} (\vec{R}; \tau) & = & \frac{1}{R} \delta(\tau - R / c) +
\lambda \frac{c \tau}{2 R^2} \delta(\tau -
R / c) \; = \; G_{2 2} (\vec{R}; \tau) \label{greena1a2} \;, \\
G_{3 0} (\vec{R}; \tau) & = & - \lambda \frac{c \tau}{2 R^2}
\delta(\tau - R / c) - \frac{\lambda \tau}{2 R} \delta'(\tau - R /
c)\;, \label{greena3}
\end{eqnarray}
where $R \equiv |\vec{R}|$, the prime on the delta function means
derivative with respect to its argument. It is interesting to
notice that the effect of the noncommutativity appears as a
$\delta'$ coming from the shifted poles in the $\omega$ integrals.
This means that the difference in the propagation speeds, $c$ and
$c'$, will be converted into a pre-acceleration effect due to
$\ddot \beta$ (see below) at one single speed of light $c$.

\noindent One can now compute the electric and magnetic fields
from the general expression for the potentials given by
\begin{equation}\label{agjxt}
A_\mu (\vec{x},t) = \frac{1}{c} \int d^3 x' d t' G_{\mu \nu}
(\vec{x} - \vec{x}'; t - t') J_\nu (\vec{x}', t') \;.
\end{equation}
>From the structure of the Green's functions in
(\ref{greenphi})-(\ref{greena3}) one can sees that
\begin{equation}\label{aeb}
A_\mu = A^{(0)}_\mu + \lambda A^{(\lambda)}_\mu \; , \; \vec{E} =
\vec{E}^{(0)} + \lambda \vec{E}^{(\lambda)} \; , \; \vec{B} =
\vec{B}^{(0)} + \lambda \vec{B}^{(\lambda)}
\end{equation}
where $A^{(0)}_3 = 0$, and $\vec{B}^{(0)} = \vec{n} \times
\vec{E}^{(0)}$.

\noindent The electric and magnetic fields have quite involved
expressions , and the part proportional to $\lambda$ contains a
term of the form \cite{ciz}
\begin{equation}\label{beta..}
  \left[\frac{1}{c (1 - \vec{n} \cdot
\vec{\beta})} \frac{d}{d t'} \left( \frac{1}{c (1 - \vec{n} \cdot
\vec{\beta})} \frac{d}{d t'} \frac{\vec{n} c (t - t')}{(1 -
\vec{n} \cdot \vec{\beta}) R} \right) \right]_{\rm ret} \;,
\end{equation}
where $\vec{n} = \vec{R} / R$, and $[\quad]_{\rm ret}$ are the
usual retarded quantities. We see here the announced contributions
proportional to the {\it derivative} of the acceleration. As
discussed earlier, the $\ddot \beta$ contribution arises as an
effect of the conversion of the two speeds  of lights $c$ and $c'$
in the poles of the Green's function into a single speed $c$ with
a {\it derivative} of the delta function $\delta'(\tau - R /c)$.
We have taken this view, rather than retaining {\it both} speeds,
in order to better compare our results with the experiments.

\noindent Let us say here that these terms, which are introduced
solely by noncommutativity, recall the familiar acausal scenario
of the Abraham-Lorentz pre-acceleration effects for the classical
self-energy of a point charge \cite{jackson}. Even if not directly
connected to the Abraham-Lorentz case, this feature is quite
surprising in this context. As a matter of fact, we naturally
obtain this effect by retaining only one space component of
$\theta^{\mu \nu}$, while it seems that one should expect such
behaviors only for non-zero time components of $\theta^{\mu \nu}$.

\section{Corrections to Synchrotron Spectrum}

\noindent In order to compare our results with the standard ones,
we want now to compute the effects of noncommutativity on the
synchrotron radiation in the experimentally relevant case defined
by the following approximations:
\begin{itemize}
  \item Ultra-relativistic motion $\beta = v / c \to 1$;
  \item Radiation observed in the plane $(1,2)$, and far from the source $|\vec{x}| \sim R >>
  |\vec{r}(t)|$.
  \end{itemize}
The power radiated in the direction $\vec n$ is
\begin{equation}
  \frac{d P (t)}{d \Omega} = R^2 [\vec{S} \cdot \vec{n}]
  \label{power12}\;,
\end{equation}
where all the quantities ($\vec{n}, \vec{\beta}$,
$\dot{\vec{\beta}}$, $R$) are in the plane $(1,2)$, and we used
the modified Poynting vector given in Eq. (\ref{poynting})
\[
\vec{S} = \frac{c}{4 \pi} \vec{D} \times \vec{B} \;.
\]
One can easily verify that in these approximations
$A^{(\lambda)}_3$ does not contribute to the radiated power in Eq.
(\ref{power12}), and for the evaluation of the order of magnitude
of the leading noncommutative correction to the power one can use
the following approximate expressions for the electric and
magnetic fields
\begin{eqnarray}
\vec{E} (\vec{x}, t) & \sim & e  \left[ \frac{1}{c \zeta}
\frac{d}{d t'} \frac{\vec{n} - \vec{\beta}}{\zeta R}
 - \lambda \; \frac{1}{c \zeta} \frac{d}{d t'} \left( \frac{1}{c \zeta}
\frac{d}{d t'}  \frac{\vec{n} c (t - t')}{\zeta R} \right)
\right]_{\rm ret} \label{e} \;, \\
\vec{B} (\vec{x}, t) & \sim & e \left[ \frac{1}{c \zeta}
\frac{d}{d t'} \frac{\vec{\beta} \times \vec{n}}{\zeta R}
\right]_{\rm ret} \label{b} \;,
\end{eqnarray}
where $\zeta \equiv 1 - \vec{n} \cdot \vec{\beta}$. The
expressions (\ref{e}) and (\ref{b}) for the electric and magnetic
fields, in the limit $\lambda =0$, reproduce the standard results
for the terms $O (1 / R)$ in the ultra-relativistic limit
\cite{jackson}, which are the only relevant ones for the
evaluation of the synchrotron radiation.

\noindent By using the expressions (\ref{e}) and (\ref{b}) for the
electric and magnetic fields, and retaining only the leading
contributions for large $R$, and $\beta \to 1$, it turns out that
\cite{ciz}
\begin{equation}\label{power2}
\frac{d}{d \Omega} P (t) \equiv |\vec{L}(t)|^2 \sim
|\left(\frac{e^2}{4 \pi c}\right)^{1/2} \left[ (1 + \frac{3
\lambda}{2 \zeta} ) \frac{1}{\zeta^3} \; \vec{n} \times (
\vec{\beta} \times \dot{\vec{\beta}}) \; \right]_{\rm ret} |^2 \;,
\end{equation}
where $\vec{L}(t) \equiv \sqrt{c / 4 \pi} [R (\vec{E}^{(0)} +
\lambda / 2 \vec{E}^{(\lambda)})]_{\rm ret}$.

\noindent The energy radiated in the plane is then \cite{jackson}
\begin{equation}\label{energy}
\frac{d}{d \Omega } I (\omega) = 2 |\vec{L}(\omega)|^2 \;,
\end{equation}
where $\vec{L} (\omega)$ is the Fourier transform of $\vec{L} (t)$
given by
\begin{equation}\label{lomega}
  \vec{L} (\omega) = \left(\frac{e^2}{8 \pi^2 c}\right)^{1/2}
  \int dt' e^{- i \omega (t' - \vec{n} \cdot \vec{r}(t'))} \vec{n} \times (\vec{n} \times \vec{\beta})
  [- i \omega (1 + \frac{3 \lambda}{2 \zeta}) + \frac{3 \lambda}{2 \zeta^3} \vec{n} \cdot
  \dot{\vec{\beta}}]\;.
\end{equation}
In the ultra-relativistic approximations two are the
characteristic frequencies for the synchrotron: the cyclotron
frequency $\omega_0 \sim c / |\vec{r}|$, and the critical
frequency $\omega_c = 3 \omega_0 \gamma^3$. In order to consider
only the radiation in the plane, we shall work in the range of
frequencies $\omega >> \omega_0$, for which the latitude
$\vartheta \sim \pi /2$ \cite{jackson}.

\noindent In this setting the leading terms for the energy
radiated in the plane are \cite{jackson}, \cite{ciz}
\begin{equation}\label{energy2}
\frac{d}{d \Omega} I (\omega) \sim \frac{e^2}{3 \pi^2 c} \left(
\frac{\omega}{\omega_0} \right)^2 \gamma^{-4} \left[ K_{2/3}^2
(\xi) [1 + \lambda (1 + 6 \gamma^2)] + \lambda \frac{24 \gamma^5
\omega_0}{\omega} K_{1/3} (\xi) K_{2/3} (\xi) \right] \;,
\end{equation}
where
\begin{equation}\label{xi}
  \xi = \frac{\omega}{ 3 \omega_0} \gamma^{-3} \;,
\end{equation}
$K_{2/3} (\xi)$, and $K_{1/3} (\xi)$ are the modified Bessel
functions. The formula in Eq. (\ref{energy2}) reproduces the
standard results in the case $\lambda =0$. When $\omega <<
\omega_c$, $\xi \to 0$, and $K_{\nu} (\xi) \sim \xi^{- \nu}$, $\nu
= 2/3 , 1/3$.

\noindent When $\omega_0 << \omega << \omega_c$, i.e. $1 << \omega
/ \omega_0 << \gamma^3$,
\begin{equation}\label{ratio}
X \equiv \frac{d I (\omega) / d \Omega}{d I(\omega) / d
\Omega|_{\lambda = 0}} \sim 1 + 10 (\frac{\omega_0}{\omega})^{2/3}
\lambda \gamma^4 \;.
\end{equation}

\noindent By using the current bound on the parameter of
noncommutativity \cite{bound} $\theta < 10^{-2} ({\rm TeV})^{-2}$,
one has that $\lambda = 2 b \theta < 2 n 10^{-23}$, where $n$ is
the value of $b$ in Tesla, and 1 Tesla $\sim 10^{-21} ({\rm
TeV})^{2}$. Thus for an electron synchrotron the correction is
\begin{equation}
X < 1 + (\frac{\omega_0}{\omega})^{2/3} n \times 10^{-21} \times
\left( \frac{{\cal E} ({\rm MeV})}{{\rm MeV }}\right)^{4} \;,
\end{equation}
where $\cal E$ is the energy of the electron, $\gamma_{\rm max}
\sim 2 {\cal E} ({\rm MeV}) / {\rm MeV}$.

\noindent For instance, for the most energetic synchrotron
(SPring-8, Japan) ${\cal E} = 8$ GeV, $b \sim 1$ Tesla, and when
$\omega / \omega_0 \sim \gamma^2$ we have
\[
X  < 1 + 10^{-10} \,.
\]
Thus, there is an impressive ''amplification'' of the effects
induced by a nonzero noncommutativity parameter (better, our
$\lambda$). As a matter of fact one gains 13 orders of magnitude
(from $10^{-23}$ to $10^{-10}$ for the case considered in this
example), and this gain is independent from the actual input value
for $\theta$.

\section{Conclusions}

\noindent We investigated synchrotron radiation in noncommutative
classical electrodynamics. The spectrum of this radiation is
considerably modified by noncommutativity of the coordinates.
These departures from the standard spectrum work as an impressive
''amplifier'' of the noncommutative effects. On the other hand we
cannot obtain a tight bound on $\theta$. This study indicates that
the phenomenology of synchrotron radiation in noncommutative
electrodynamics deserves further investigation.

\noindent  We notice a partial analogy of the present theory, 
in the linearized approximation,  with nonlinear optic. A comparison 
of the latter  with the noncommutative case could give interesting results
because the properties of the "medium" (as described by  $\epsilon _{ij}$ 
and  $\mu _{ij}$), depend  in our framework on the external background 
magnetic field $\vec b$. It is then possible to separate the effects
of noncommutativity, coupled to $\vec b$, from the other 
electrodynamical effects.

\noindent We also saw a peculiar acausal behaviour for the
electric and magnetic fields as time-derivatives higher than two.
This Abraham-Lorentz-like effect, which naturally arose even if
only one space-space component of $\theta^{\mu \nu}$ is taken to
be different from zero, is due to the two different speeds of
light allowed in this theory. Moreover, it is in contrast with the
general belief that such acausal effects should arise in this
context only when $\theta^{0 i} \neq 0$. We take this last result
as a confirmation that the fascinating issues of the space-time
properties in presence of noncommuting coordinates is far from
being fully clarified.

\vspace{.5cm}

\noindent {\bf \large Acknowledgments}

\noindent We thank Roman Jackiw for useful discussions. A.I.
thanks Victor Rivelles for advise regarding the literature. 
This work is supported in part by funds provided by the U.S.
Department of Energy (D.O.E.) under cooperative research agreement
DF-FC02-94ER40818.



\end{document}